\documentclass[final,twocolumn,abstractlogo]{ise}
\usepackage[switch]{lineno}
\usepackage{cuted}
\usepackage{wrapfig}
\newcommand{\toolname}{\textsc{NanoHarness}\xspace}
\usepackage{mdframed}
\usepackage{bbding}

\usepackage{tcolorbox}
\tcbuselibrary{breakable}
\usepackage{pifont}
\usepackage{enumerate}
\usepackage{ragged2e}
\usepackage{makecell}
\usepackage{stfloats}
\usepackage[vlined, ruled]{algorithm2e}
\usepackage{algorithmic}
\usepackage{textcomp}
\usepackage{multirow}
\usepackage{diagbox}
\usepackage{listings}
\usepackage{float}
\usepackage{url}
\usepackage{xspace}
\usepackage{booktabs}
\usepackage{multirow}
\usepackage{makecell}
\usepackage{siunitx}
\usepackage{xcolor}
\usepackage{tabularx}
\usepackage{calc}

\definecolor{deepblue}{rgb}{0,0,0.5}
\definecolor{deepgreen}{rgb}{0,0.5,0}
\definecolor{deepred}{rgb}{0.6,0,0}
\definecolor{darkorange}{RGB}{255,140,0}
\definecolor{lightgray}{rgb}{0.93,0.93,0.93}
\definecolor{deepgray}{rgb}{0.25,0.25,0.25}

\tcbuselibrary{skins}
\lstdefinelanguage{Java}{
	basicstyle=\small\ttfamily,
	numberstyle=\color{deepgray},
	stepnumber=1,
	numbersep=8pt,
	showstringspaces=false,
	breaklines=true,
	frame=lines,
	backgroundcolor=\color{lightgray},
	commentstyle=\color{deepgreen},
	keywordstyle=\color{deepblue},
	stringstyle=\color{deepred},
	tabsize=4,
	captionpos=b,
	morekeywords={public, class, void, int, if, else, for, while, return, true, false},
	emph={String, System},
	emphstyle=\color{darkorange},
	alsoletter={.,;:[]()},
}
\usepackage{color}
\usepackage{stmaryrd}
\usepackage{verbatimbox}
\usepackage{makecell}
\usepackage{booktabs}

\usepackage{colortbl}

\newcommand{\finding}[2]{
\begin{center}
\begin{tcolorbox}[leftrule=0mm,toprule=0mm,bottomrule=0mm,rightrule=0mm,left=1pt,right=2pt,top=0pt,bottom=0pt,breakable]
\textbf{Answer to RQ{#1}:}
{#2}
\end{tcolorbox}
\end{center}
}
\usepackage[normalem]{ulem}

\newcommand{\delete}[1]{\iffalse{#1}\fi}

\AtBeginDocument{%
  \providecommand\BibTeX{{%
    \normalfont B\kern-0.5em{\scshape i\kern-0.25em b}\kern-0.8em\TeX}}}
    
\title{Beyond the Model: Demystifying Harness Effects in Software Engineering Agents}
\shorttitle{Measuring Harness Effects in Software Engineering Agents}

\newcommand{\authorname}[1]{{\sffamily\normalsize\bfseries\color{ISEInk}#1}}

\author{%
  \authorname{Haichuan Hu\textsuperscript{1}} \quad
  \authorname{Quanjun Zhang\textsuperscript{1}} \quad
  \authorname{Shengcheng Yu\textsuperscript{2}} \quad
  \authorname{Zhifei Chen\textsuperscript{1}} \\[0.3mm]
  \authorname{Tianyu Luo\textsuperscript{3}} \quad
  \authorname{Chunrong Fang\textsuperscript{3}} \quad
  \authorname{Zhenyu Chen\textsuperscript{3}} \quad
  \authorname{Liang Xiao\textsuperscript{1}}
}
\affil{%
  {\normalfont\normalsize\color{black}
    \textsuperscript{1}Nanjing University of Science and Technology \quad
    \textsuperscript{2}Technical University of Munich\\[-0.1ex]
    \textsuperscript{3}Nanjing University}\\[1.1mm]
  {\normalfont\footnotesize\ttfamily\color{black}
    huhaichuan2024@gmail.com, quanjunzhang@njust.edu.cn, shengcheng.yu@tum.de\\[-0.1ex]
    chenzhifei@njust.edu.cn, Ty\_L191025@outlook.com, fangchunrong@nju.edu.cn\\[-0.1ex]
    zychen@nju.edu.cn, xiaoliang@mail.njust.edu.cn}
}

\keywords{Large Language Models, Agent Harness, Software Engineering, Code Generation, Coding Agents}
\date{September 26, 2026}

\begin{document}

\maketitle

\begin{abstract}
Large Language Model (LLM)-based agents are increasingly used for software engineering tasks, yet their performance is not determined by the base model alone. The agent harness substantially shapes how SE agents interact with repositories, execute actions, and validate solutions. However, the role of harness design remains insufficiently understood, especially across different models, tasks, and harness components. In this paper, we present a systematic empirical study of harness effects in SE agents. We first evaluate two representative harnesses, mini-SWE-agent and OpenCode, with ten models from two prominent open-weight model families, Qwen and DeepSeek, on three benchmarks: SWE-bench Pro, ProgramBench, and GitTaskBench. We then construct \toolname{}, a lightweight modular harness built on top of mini-SWE-agent, and use it to analyze five representative harness components: tool registry, context compression, explicit planning, subagents, and lazy skills. 
Experimental results show that harness effectiveness depends jointly on model capability and task type. Complex harnesses provide diminishing marginal gains on SWE-style issue repair as model capability improves, but can benefit stronger models on more complex and open-ended repository-level tasks. Component-level analysis on ProgramBench further shows that structured tool use and task-specific subagents provide the most stable improvements, while context compression and general subagents can hurt repository-generation performance. When combined, \toolname{} improves over mini-SWE-agent by 7.37 and 6.21 percentage points on Qwen3.7-Max and DeepSeek-V4-Pro, respectively, recovering most of the gains of product-level harnesses. 
These findings highlight harness design as a first-class factor in SE-agent performance and provide insights for building more effective and efficient coding agents.
\end{abstract}

\section{Introduction}
Large Language Model (LLM)-based agents are increasingly used to solve Software Engineering (SE) tasks, such as issue resolution~\cite{deng2025swe,yang2024swe,Zhang2026SGAgentSL}, code modification~\cite{fan2025exploring,huang2025comprehensive,li2025hybrid}, and test generation~\cite{lops2025system,wang2025testeval}. While much progress has been driven by stronger foundation models (e.g., GPT~\cite{singh2025openai,achiam2023gpt}, Claude~\cite{anthropic2024claude3modelcard}, Gemini~\cite{team2023gemini}, DeepSeek~\cite{liu2024deepseek}, Qwen~\cite{yang2025qwen3}), the performance of coding agents also depends heavily on the model-external infrastructure that surrounds the model, including context retrieval~\cite{ouyang2025repograph,zhang2023repocoder}, tool use~\cite{ding2025toolcoder,zhang2024codeagent}, execution environments~\cite{xi2025agentgym}, patch generation~\cite{shao2026fix,kim2025logs}, and validation~\cite{dong2025codescore,zhang2026compass}. This infrastructure is commonly referred to as the agent harness~\cite{meng2026agent,zhou2026externalization}.

\begin{figure*}[!t]
\centering
\includegraphics[width=\textwidth]{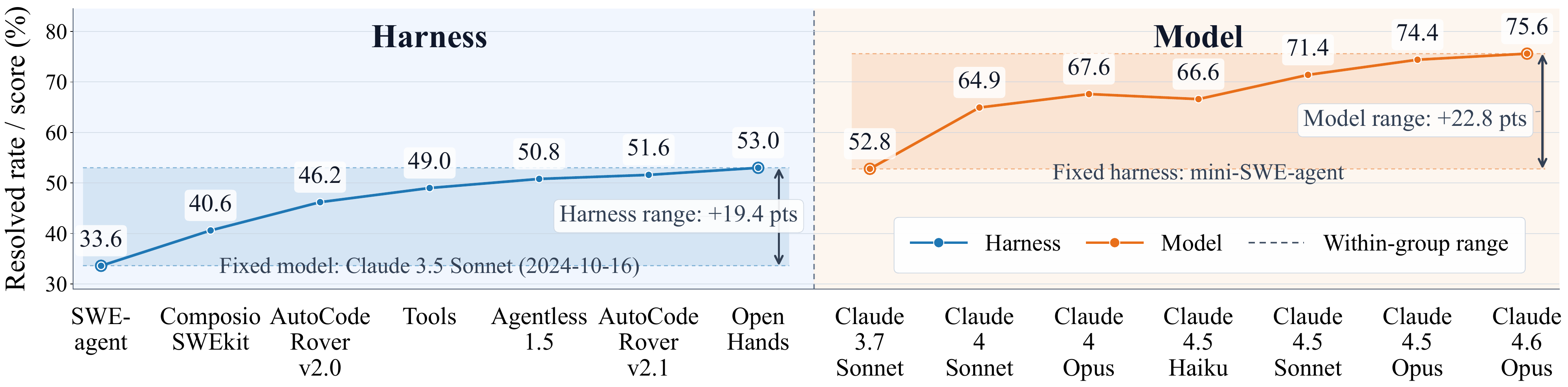}
\caption{Comparing fixed-model and fixed-harness settings on SWE-bench Verified. Results are collected from the leaderboard.}
\label{fig:harness_vs_model}
\end{figure*}

Publicly reported results on the SWE-bench Verified leaderboard~\cite{jimenez2024swe} suggest that harness design can have an impact comparable to model choice. As shown in Figure~\ref{fig:harness_vs_model}, the gap between the best- and worst-performing harnesses under a fixed model reaches 19.4 percentage points, close to the 22.8-point gap between Claude models under a fixed harness. Epoch AI~\cite{epoch2025whatskillsdoesswebenchverifiedevaluate} also points out that harnesses matter as much as model choice, and that a well-designed harness can yield roughly a 20\% performance improvement.
These motivates a controlled analysis of which harness design choices drive such gains and whether they transfer across models and tasks.

Despite this impact, existing SE research has not yet isolated harness design as an object of systematic study. Recent work~\cite{wang2025swe,li2025swe,bouzenia2025repairagent,yu2025patchagent,wang2025aegis} has introduced increasingly capable coding agents for specific software engineering scenarios. These agents necessarily make concrete harness-level choices, such as how to retrieve repository context, expose tools, plan edits, run tests, and select final patches. However, these choices are typically evaluated as part of an end-to-end agent system, rather than analyzed as separable design components. As a result, we still lack a clear understanding of which harness components drive improvements, how they interact, and how their effects depend on base models and task types.

To bridge this gap, we conduct an empirical study of harnesses in SE agents. Our study focuses on two complementary perspectives: 
(1) how models, harnesses, and tasks jointly shape agent performance; and (2) how harness components contribute to performance on general SE tasks.

For the first perspective, we evaluate agents on three representative SE tasks that broadly cover the SE lifecycle: issue repair on SWE-bench Pro~\cite{deng2025swe}, repository-level code generation on ProgramBench~\cite{yang2026programbench}, and repository-centric real-world task solving on GitTaskBench~\cite{ni2026gittaskbench}. 
We study these tasks under two contrasting harness configurations: mini-SWE-agent~\cite{yang2024swe}, a lightweight harness with basic command-line interaction capabilities, and OpenCode~\cite{opencode2026}, a product-level harness with a more complex design. 
We pair these harnesses with two model families, Qwen~\cite{yang2025qwen3} and DeepSeek~\cite{liu2024deepseek}, selecting five representative models from each family. 
Overall, this design yields 60 experimental configurations (3 tasks $\times$ 2 harnesses $\times$ 10 models), enabling a systematic analysis of how models, harnesses, and tasks jointly shape agent performance.

For the second perspective, we follow commercial harness designs (e.g., Claude Code~\cite{anthropic_claude_code} and Codex~\cite{openai_codex}) to identify five core components of modern harness architectures, and implement each component in a minimal form.
Starting from mini-SWE-agent, we incrementally add these plug-and-play components to the existing harness in a building-block manner, creating a controlled harness variant called \toolname{} for component-level analysis.
Using ProgramBench as the testbed, we evaluate both the isolated effect of each component and the combined effect of integrating all components.
In addition, we collect and analyze both the performance and execution behavior of \toolname{} and its variants, and further conduct trajectory-level analysis.

Experimental results show that harness design substantially shapes SE-agent performance, but its effect is neither uniform nor monotonic. First, harnesses play different roles under different task regimes. On SWE-style issue repair, where workflows are relatively fixed and edits are often localized, complex harnesses mainly compensate for weaker models, and their marginal benefit decreases as model capability improves. In contrast, on newer and more open-ended repository-level tasks such as ProgramBench and GitTaskBench, complex harnesses behave more like capability amplifiers: only sufficiently capable models can reliably exploit the additional workflow complexity.

Second, our component and trajectory analyses show that harness gains come less from adding more scaffolding or spending more tokens, and more from regulating how agents explore and act in repositories. Structured tool use and task-specific subagents provide the most stable improvements, while context compression and generic subagents can hurt repository-generation performance. \toolname{} improves over mini-SWE-agent by 7.37 and 6.21 percentage points on Qwen3.7-Max and DeepSeek-V4-Pro, respectively, and closes most of the gap to product-level harnesses with only controlled prompt growth. These results suggest that effective harnesses improve SE agents by turning raw exploration into structured, task-aligned interaction rather than by increasing model capability alone.

This paper makes the following contributions: 
(1) \textbf{Large-scale Empirical Study.} 
We conduct a systematic empirical study across three representative benchmarks, using 10 different LLMs from Qwen and DeepSeek, with over \textbf{10B} (Qwen) and \textbf{20B} (DeepSeek) prompt tokens. 
(2) \textbf{Component-Level Decomposition.} 
We decompose modern SE-agent harnesses into representative components and build \toolname{}, a lightweight modular harness that enables controlled component-level analysis. 
(3) \textbf{Behavioral Insight.} 
We reveal how harness mechanisms affect performance and execution behavior, showing that effective harnesses improve SE-agent performance mainly by enabling structured tool use and better-regulated exploration rather than by model improvement alone.

\section{Preliminaries} \label{sec:preliminary}
In this section, we briefly introduce the architecture and main components of the modern harness design. 

\subsection{Architecture}
\begin{figure*}[!t]
\centering
\includegraphics[width=\textwidth]{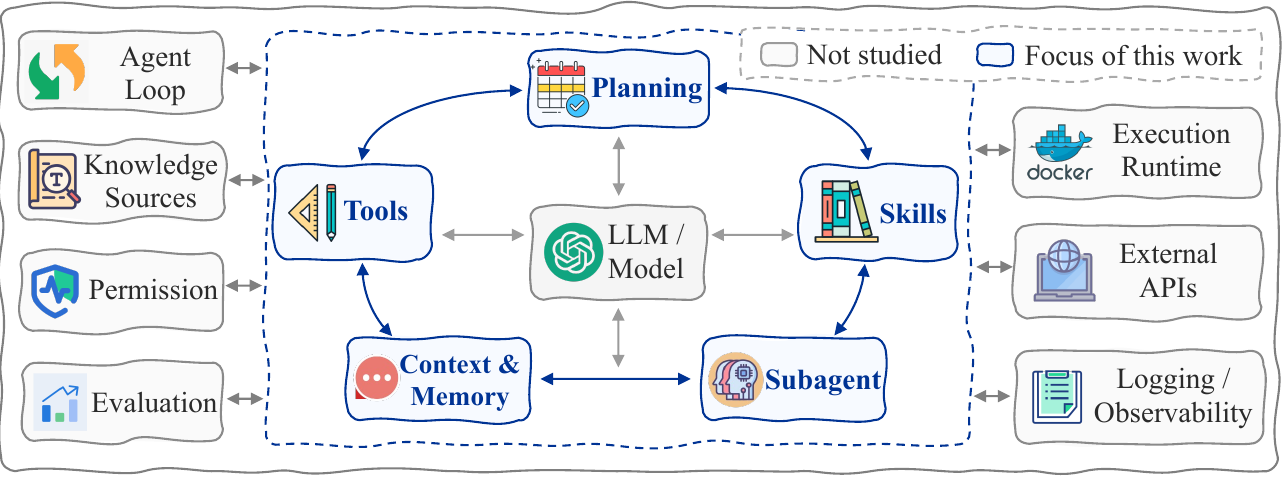}
\caption{ The general architecture of modern agent harnesses. }
\label{fig:arch}
\end{figure*}

As shown in Figure~\ref{fig:arch}, we present a general architecture of modern agent harnesses. 
The harness is organized around the model, and we decompose it into 12 modules~\cite{meng2026agent,zhou2026externalization}. The modules highlighted in blue are the focus of this work, as they represent the fundamental capabilities of a harness and have a direct impact on agent performance. The gray modules include supporting components such as environment, permission, logging, observability, knowledge sources, and evaluation. Although these modules are also indispensable parts of a complete agent system, they usually play more engineering-oriented roles or serve as external constraints across different harnesses. Therefore, we do not discuss their internal mechanisms in detail. Next, we introduce the five core modules of the harness.

\subsection{Components}
\subsubsection{Tools}
Tools expose external actions such as file inspection, editing, search, shell execution, and testing. Their interface determines what repository state the model can observe and how safely it can act on that state.

\subsubsection{Context and Memory}
Context and memory determine what task information, observations, code snippets, outputs, and intermediate reasoning remain available during the agent loop. Effective management preserves useful state while limiting irrelevant or redundant history.

\subsubsection{Planning}
Planning helps the agent decompose a task, track intermediate goals, and decide subsequent actions. Harnesses may implement this through explicit plans, iterative replanning, or lightweight action selection.

\subsubsection{Subagents}
Subagents delegate parts of a task to specialized roles or execution flows, such as localization, implementation, or verification. This can improve decomposition but also adds coordination and communication overhead.

\subsubsection{Skills}
Skills encode reusable high-level workflows such as debugging, validation, or recovery from failed attempts. They reduce repetitive decision-making when the procedure is relevant to the current task.

\section{Study on SE Agent Harness}
\subsection{Overall Study Design}

\begin{figure*}[!t]
\centering
\includegraphics[width=\textwidth]{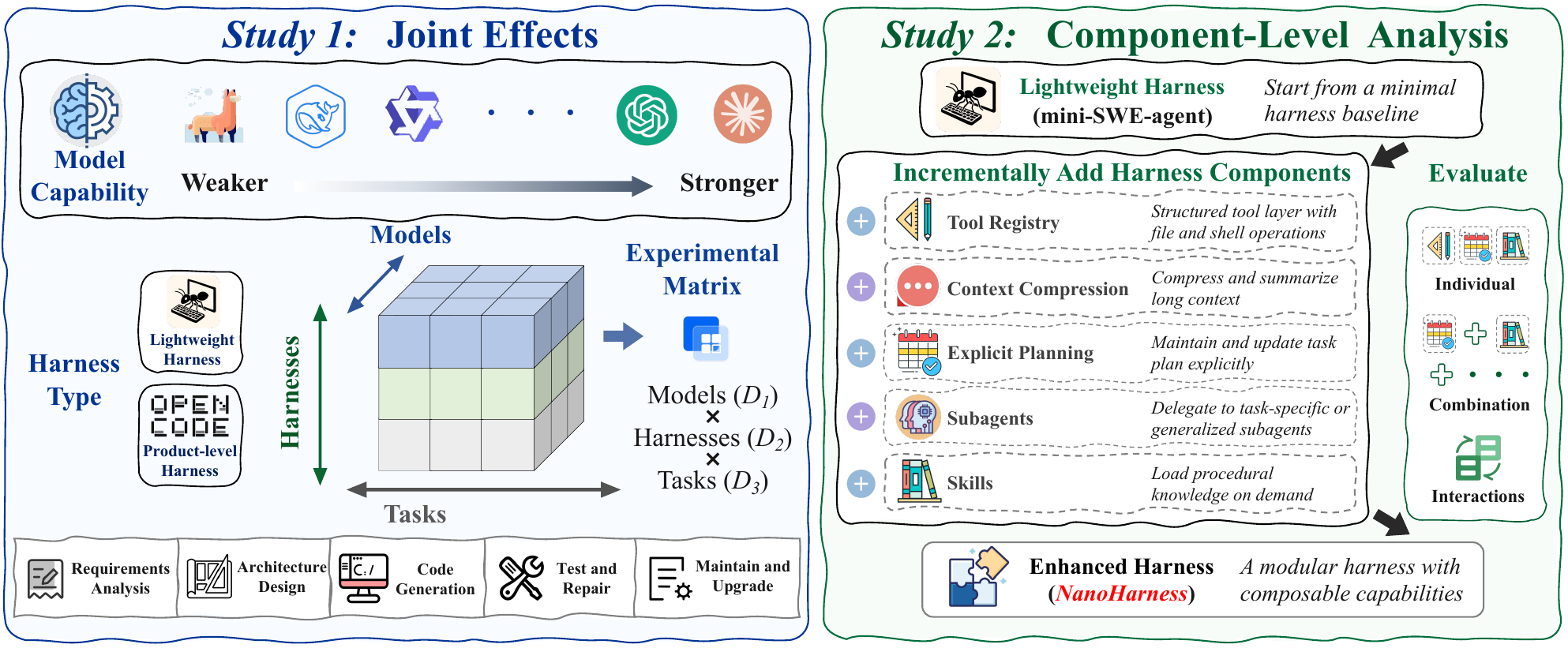}
\caption{Overview of our agent harness study: (1) joint model–harness–task effects and (2) component-level harness analysis.}
\label{fig:study}
\end{figure*}

\textbf{Overview.} Figure~\ref{fig:study} provides an overview of our empirical study on SE agent harnesses. The study is organized into two complementary parts. First, we examine the joint effects of models, harnesses, and tasks by evaluating models with different capability levels under both lightweight and product-level harnesses across SE tasks spanning the software development lifecycle. Second, we conduct a component-level analysis by incrementally adding core harness mechanisms, including tool registry, context compression, explicit planning, subagents, and lazy skills, to a minimal baseline harness. This design allows us to study not only how harnesses interact with model capability and task characteristics, but also how individual components and their combinations affect agent performance, behavior, and efficiency.

\textbf{Study 1: Measuring Joint Effects of Models, Harnesses, and Tasks.} SE research mainly follows three directions: improving coding-oriented models, designing agent harnesses, and building SE tasks and benchmarks. Model-oriented studies~\cite{Shang2024ALE,Yu2025SmartLLaMADPORL,Yang2024MORepairTL} train models through code pretraining, fine-tuning, reinforcement learning, or inference-time optimization. Harness-oriented studies~\cite{Xia2024AgentlessDL,Bouzenia2024YouNI,Wang2025AEGISAA} improve how agents use repositories, tools, tests, and execution environments. Task-oriented and benchmark-oriented studies~\cite{Shafin2025EvaluatingSP,Wang2025Defects4CBL} design realistic SE tasks and evaluation protocols. However, these directions are usually studied with separate focuses, leaving the joint effects of models, harnesses, and tasks underexplored.

To examine these joint effects, we design our first study around the three factors. For models, we select multiple models from the same model family, ranging from weaker to stronger ones, so that performance differences can reflect changes in model scale and capability while reducing variation caused by model lineage. For harnesses, we compare a lightweight basic harness with a more complex product-level harness, which allows us to measure how much the surrounding agent workflow changes SE task performance. For tasks, we choose benchmarks that cover different stages of the software development lifecycle, including requirements analysis~\cite{demirel2018software}, architecture design~\cite{zhang2025knowledge}, coding~\cite{jiang2026survey}, testing~\cite{lops2025system}, and validation~\cite{dong2025codescore}. By combining these factors in a unified three-dimensional experimental matrix, we evaluate each model under each harness across each task, allowing us to compare their individual effects and analyze how they interact in LLM-based SE agents.

\textbf{Study 2: Component-Level Harness Analysis.}
The second study focuses on the internal design of the agent harness. As discussed in Section~\ref{sec:preliminary}, modern SE agent harnesses can be decomposed into 12 components. To study the mechanisms behind representative components, we implement a modular harness called \toolname{} on top of mini-SWE-agent, where each mechanism can be enabled or disabled independently. Specifically, we instantiate five representative mechanisms: tool registry, context compression, explicit planning, subagents, and lazy skills. This modular design allows us to evaluate their effects on SE tasks both individually and in combination, thereby analyzing how different harness mechanisms contribute to long-horizon performance beyond the choice of model alone.

\textbf{Tool Registry.} We extend the Bash interactive mode of mini-SWE-agent into a tool layer, implemented through a unified registry and native tool-calling mechanism. We introduce four file-system-oriented tools: \texttt{read\_file}, \texttt{write\_file}, \texttt{edit\_file}, and \texttt{glob}. \texttt{read\_file} provides bounded file inspection, \texttt{write\_file} creates or overwrites files, \texttt{edit\_file} applies localized textual modifications, and \texttt{glob} supports pattern-based file discovery. 

\textbf{Context Compression.} We implement a context compression module for long-horizon execution under limited context windows. Oversized tool outputs are persisted to the workspace, with only paths and short previews retained in the prompt. When history grows too long, the module applies message snipping and observation compaction while preserving recent results and high-value outputs. If the context still exceeds the limit, it triggers LLM-based summarization to retain necessary information. Compaction can also be invoked manually or after context-window errors, with logs kept for auditing.

\textbf{Explicit Planning.} The planning module adds a TodoWrite-style lightweight planning mechanism to the agent. When enabled, the agent can call \texttt{todo\_write} to maintain a short task list, where at most one item may be marked as \texttt{in\_progress}. Each update replaces the full todo list and is validated by \texttt{TodoManager} for item format, status values, list length, and planning consistency. The current plan is rendered back into the prompt and saved in the trajectory metadata. To avoid stale plans, the agent injects reminders when the plan has not been updated for several rounds.

\textbf{General/Task-Specific Subagents.}
The subagent module provides on-demand specialist capabilities while leaving task control, budget management, and final submission to the parent agent. Each subagent runs with fresh history in the shared workspace, uses a restricted tool whitelist via \texttt{SubagentRunner}, and returns a structured report through \texttt{finish\_subagent}. Nested delegation is disabled, and trajectories are logged for auditing and cost analysis.

\noindent\textbf{(1) General SE subagents.}
The first type consists of software-engineering subagents organized according to the software development lifecycle, including \textit{requirement\_analysis}, \textit{architecture\_design}, \textit{code\_generation}, \textit{test\_generation}, and \textit{defect\_repair}. Each subagent is assigned an appropriate tool set and returns structured artifacts for the parent agent to integrate.

\noindent\textbf{(2) Task-specific subagents.}
The second type consists of dedicated profiles for \textit{behavior analysis}, \textit{differential testing}, and \textit{submission review}. These subagents support targeted exploration, validation, and final checking.

\textbf{Lazy Skills.}
The skills module reduces prompt overhead by treating procedural knowledge as lazily loaded, callable resources. When enabled, the prompt contains only a compact skill catalog, and the agent can invoke \texttt{load\_skill(name)} to retrieve detailed guidance when needed. \texttt{SkillLoader} builds this catalog from metadata in configured \texttt{SKILL.md} files. Loaded skills are returned in a structured wrapper and support implementation, debugging, testing, fuzzing, and final review. The available skills are defined in Table~\ref{tab:skills}.

\begin{table}[h]
\centering
\caption{Overview of lazily loaded skills.}
\label{tab:skills}
\small
\renewcommand{\arraystretch}{1.06}
\begin{tabularx}{\linewidth}{@{}>{\raggedright\arraybackslash}p{0.30\linewidth}>{\raggedright\arraybackslash}X@{}}
\toprule
\textbf{Skill} & \textbf{Purpose} \\
\midrule
\textit{Terminal Workflow} & Guides terminal-based task solving from exploration to validation and submission. \\
\textit{Minimal Implementation} & Encourages small, targeted implementations and rapid validation. \\
\textit{Build Diagnostics} & Helps diagnose build, executable, toolchain, and dependency issues. \\
\textit{Test Triage} & Analyzes failed tests, logs, exit codes, and interface mismatches. \\
\textit{CLI Fuzzing} & Generates bounded edge-case tests and compares CLI behavior. \\
\textit{Submission Review} & Checks build reproducibility, executable correctness, and repository cleanliness. \\
\bottomrule
\end{tabularx}
\end{table}

\subsection{Research Questions}
Based on our study objectives, we propose the following three research questions.

\noindent\textbf{RQ1: }{What is the relationship among models, harnesses, and tasks?}
\begin{itemize}[label={}, leftmargin=22.5pt]
    \item \textbf{RQ1.1:} How does harness performance scale with model capability?
    \item \textbf{RQ1.2:} How do harnesses vary across SE tasks?
\end{itemize}

\noindent\textbf{RQ2: }{How do harness components affect SE agents?}
\begin{itemize}[label={}, leftmargin=22.5pt]
\item\textbf{RQ2.1:} What is the individual effect of each component?
\item\textbf{RQ2.2:} What is the combined effect of all components?
\end{itemize}

\noindent\textbf{RQ3: }{How do harness mechanisms affect agent execution behavior and efficiency?}
\begin{itemize}[label={}, leftmargin=22.5pt]
\item\textbf{RQ3.1:} How do harness mechanisms change the agent's tool use, context usage, and token usage?
\item\textbf{RQ3.2:} How do harness mechanisms affect common failure modes in long-horizon SE tasks?
\end{itemize}

\section{Experimental Setup}
\subsection{Tasks and Datasets}

We evaluate harnesses on three complementary SE benchmarks (Table~\ref{tab:datasets}): ProgramBench~\cite{yang2026programbench}, SWE-bench Pro~\cite{deng2025swe}, and GitTaskBench~\cite{ni2026gittaskbench}. These datasets span key stages of the SE lifecycle, including requirement understanding, architecture design, code generation, testing, and validation. This makes them suitable for assessing harness effectiveness under complex, long-horizon SE tasks. Specifically, ProgramBench targets repository-level code generation, SWE-bench Pro focuses on real-world issue repair, and GitTaskBench evaluates repository-centric task solving.

\begin{table}[t]
\centering
\caption{Summary of datasets used in our experiments.}
\label{tab:datasets}
\small
\resizebox{\linewidth}{!}{\begin{tabular}{@{}llll@{}}
\toprule
\textbf{Dataset} & \textbf{Task} & \textbf{Language} & \textbf{Size} \tabularnewline
\midrule
ProgramBench~\cite{yang2026programbench} & Repository generation & Rust, Go, C/C++, Java, Haskell & 200 \tabularnewline
SWE-bench Pro~\cite{deng2025swe} & Issue repair & Python, JS/TS, Go & 731 \tabularnewline
GitTaskBench~\cite{ni2026gittaskbench} & Repository task solving & Python & 54 \tabularnewline
\bottomrule
\end{tabular}}
\end{table}

\subsection{Harness Selection}

We select mini-SWE-agent~\cite{yang2024swe} and OpenCode~\cite{opencode2026} as two representative harnesses. mini-SWE-agent serves as the minimal harness setting, offering a concise and basic agent workflow centered on command-line interaction. It represents a lightweight harness with limited model-external support. OpenCode, in contrast, represents a product-level harness with a more complex design and multiple integrated components for repository interaction, tool use, context handling, and task execution. This contrast enables us to study how harness design complexity influences performance across different models and SE tasks.

\subsection{Model Selection}
\begin{figure}[!t]
\centering
\includegraphics[width=\linewidth]{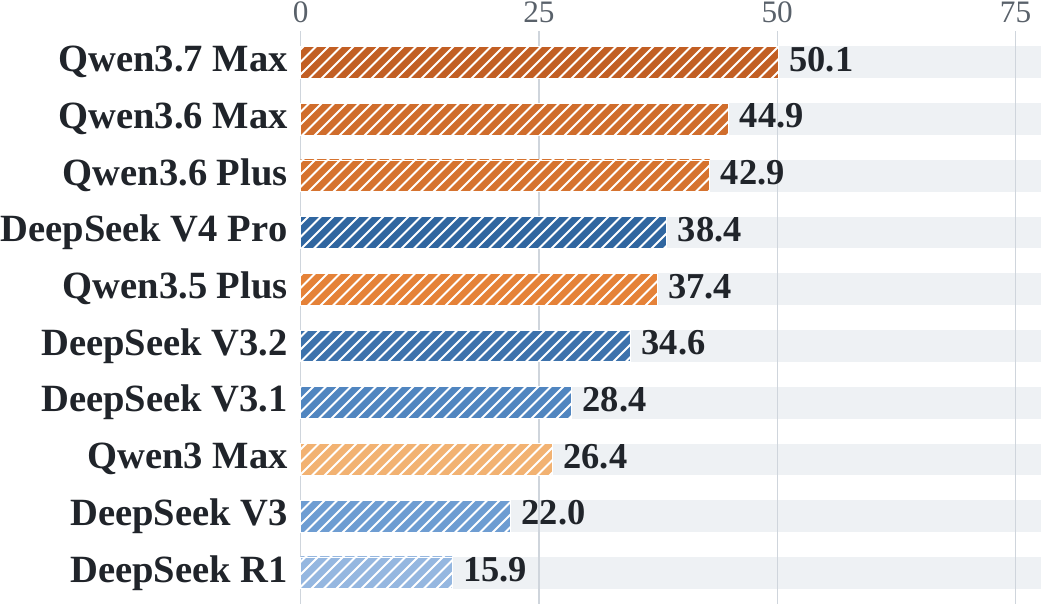}
\caption{Model coding rankings based on the Coding Index.}
\label{fig:model_rank}
\end{figure}

We focus on open-weight model families to support reproducible evaluation and controlled comparisons across models with different capability levels but similar lineage. Therefore, instead of using closed-weight models (e.g., GPT, Claude), we select two widely used open-weight model families, Qwen~\cite{bai2023qwen,yang2025qwen3} and DeepSeek~\cite{liu2024deepseek}, and choose five models from each family.
We rank the models by strength based on the official scores provided by the Coding Index~\cite{artificialanalysis_codingindex_2026} (Figure~\ref{fig:model_rank}). This selection and ranking provide an overview of the models' coding capabilities and facilitate comparisons between stronger and weaker models. Here, DeepSeek-R1 refers to the original DeepSeek-R1 release, rather than the later DeepSeek-R1-0528 update, while DeepSeek-V3 refers to DeepSeek-V3-0324. All models are evaluated in non-thinking mode.

\subsection{Implementations}
We use a unified execution protocol across all experiments. For model inference, we set the temperature to 0.0 for all models to reduce randomness. For SWE-bench Pro and GitTaskBench, we follow the original evaluation settings and limit each task to 250 agent steps. For ProgramBench, we reduce the step limit from the original 1,000 steps to 300 steps to reduce evaluation cost. Each task is also subject to a 6-hour wall-clock limit and a 3-minute timeout for each command execution.
To control evaluation cost in RQ1, we use random subsets for the larger benchmarks: RQ1.1 uses a random sample of 300 instances from SWE-bench Pro, and RQ1.2 uses a random sample of 70 instances from ProgramBench. We use the same sampled instances across all matched model-harness configurations. RQ1.2 uses all 54 GitTaskBench instances. The component-level experiments in RQ2 and RQ3 use the full 200-instance ProgramBench set.
SWE-bench Pro and ProgramBench are evaluated in Docker containers, while GitTaskBench is evaluated in a local environment. For containerized tasks, we disable network access with \texttt{--network none}, allocate 20 CPUs and 60 GB memory per container, set a 7-hour container lifetime, and disable \texttt{SYS\_PTRACE}. For GitTaskBench, we isolate the working directory and block operations outside the workspace.
For \toolname{}, we set the maximum number of subagent rounds to 50 and limit the planning module to at most 10 todo items. Context compression is enabled with a 350K-token budget, and the most recent 20 messages are preserved when compression is triggered.

\section{Results and Analysis}

\subsection{RQ1: Relationship Among Models, Harnesses, and Tasks}
\subsubsection{RQ1.1-Scaling of Harness Performance with Model Capability}

To study how harness-induced gains scale with model capability, we first analyze matched-model results from the SWE-bench Verified~\cite{jimenez2024swe} leaderboard. As shown in Table~\ref{tab:matched-model-harness-improvement}, more advanced harnesses provide larger improvements over mini-SWE-agent for weaker models, while the margin becomes smaller for stronger models. For example, the gain of OpenHands decreases from +13.60 pp with Claude 3.7 Sonnet to +5.47 pp with Claude 4 Sonnet, and the gain of SWE-agent decreases from +9.60 pp to +4.07 pp under the same model progression. This suggests that harness complexity can substantially improve weaker or mid-level models, but its marginal benefit may be compressed as model capability increases.

\begin{table}[h]
\centering
\caption{Matched-model harness improvements over mini-SWE-agent on SWE-bench Verified.}
\label{tab:matched-model-harness-improvement}
\resizebox{\linewidth}{!}{%
\begin{tabular}{llcc}
\toprule
Harness & Model (weak $\to$ strong) & mini-SWE & Gain \\
\midrule
SWE-agent & Claude 3.7 Sonnet & 52.80 & +9.60 \\
SWE-agent & Claude 4 Sonnet & 64.93 & +4.07 \\
\midrule
OpenHands & Claude 3.7 Sonnet & 52.80 & +13.60 \\
OpenHands & Claude 4 Sonnet & 64.93 & +5.47 \\
\midrule
Tools & Claude 3.7 Sonnet & 52.80 & +10.40 \\
Tools & Claude 4 Sonnet & 64.93 & +7.47 \\
Tools & Claude 4 Opus & 67.60 & +5.60 \\
\bottomrule
\end{tabular}}
\end{table}

Our controlled experiments on a random 300-instance sample from SWE-bench Pro~\cite{deng2025swe} further support this trend, but also reveal that the scaling pattern differs across model families. For the Qwen family, the effect is relatively smooth. As model capability increases from Qwen3-Max to Qwen3.7-Max, both mini-SWE-agent and OpenCode improve, while the performance gap between them gradually narrows. Specifically, the OpenCode advantage decreases from about 6 pp for Qwen3-Max to 2.67 pp for Qwen3.7-Max, indicating that stronger Qwen models rely less on the additional scaffolding provided by the product-level harness.

The trajectory statistics in Table~\ref{tab:trajectory-stats-swe-qwen} provide further evidence for this interpretation. OpenCode keeps the number of agent steps relatively stable across Qwen models, while its number of tool calls increases from 39.87 to 72.23. This suggests that stronger Qwen models can use OpenCode to perform denser tool-assisted exploration within a similar number of decision steps. In contrast, mini-SWE-agent is constrained to roughly one tool call per step, and its step count and token usage first increase and then decrease as model capability improves. This indicates that stronger Qwen models gradually become more selective under the lightweight harness, focusing on fewer but more useful interactions.

\begin{figure}[!t]
\centering
\includegraphics[width=\linewidth]{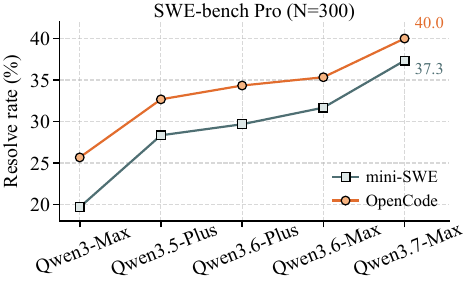}
\caption{Qwen-model scaling comparison between different harnesses on SWE-bench Pro (N=300).}
\label{fig:swe_pro_qwen}
\end{figure}

\begin{table}[t]
\centering
\caption{Average sample-level statistics of Qwen models.}
\label{tab:trajectory-stats-swe-qwen}
\resizebox{\linewidth}{!}{\begin{tabular}{llcccc}
\toprule
Harness & Model & Step & Tool & Prompt & Response \\
\midrule
mini-SWE & Qwen3-Max & 31.12 & 30.02 & 369,443 & 5,110 \\
mini-SWE & Qwen3.5-Plus & 89.72 & 88.05 & 2,347,080 & 30,942 \\
mini-SWE & Qwen3.6-Plus & 57.81 & 56.10 & 1,573,644 & 66,793 \\
mini-SWE & Qwen3.6-Max & 54.15 & 52.47 & 1,379,883 & 27,237 \\
mini-SWE & Qwen3.7-Max & 43.58 & 42.36 & 770,319 & 18,732 \\
\midrule
OpenCode & Qwen3-Max & 35.95 & 39.87 & 1,069,425 & 6,442 \\
OpenCode & Qwen3.5-Plus & 37.84 & 57.13 & 1,702,800 & 9,505 \\
OpenCode & Qwen3.6-Plus & 37.91 & 61.15 & 1,499,056 & 11,132 \\
OpenCode & Qwen3.6-Max & 36.49 & 64.87 & 1,395,815 & 11,283 \\
OpenCode & Qwen3.7-Max & 37.38 & 72.23 & 1,523,661 & 14,921 \\
\bottomrule
\end{tabular}}
\end{table}

\begin{figure}[!t]
\centering
\includegraphics[width=\linewidth]{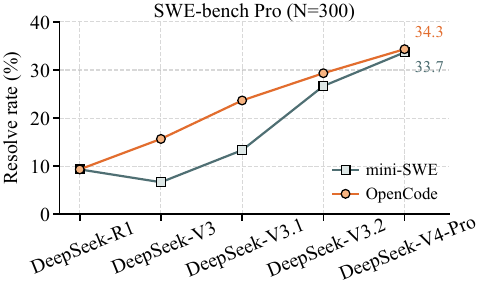}
\caption{DeepSeek-model scaling comparison between different harnesses on SWE-bench Pro (N=300).}
\label{fig:swe_pro_ds}
\end{figure}

DeepSeek exhibits a different scaling pattern, as shown in Figure~\ref{fig:swe_pro_ds}. Unlike Qwen, the harness gain does not decrease smoothly across the whole model sequence. OpenCode brings little improvement for DeepSeek-R1, larger gains for intermediate models such as DeepSeek-V3 and DeepSeek-V3.1, and then converges with mini-SWE-agent again for DeepSeek-V4-Pro, where the final scores are 34.3 for OpenCode and 33.7 for mini-SWE-agent. This non-monotonic pattern suggests a threshold effect: models that are too weak may not reliably exploit a complex harness, while sufficiently strong models can solve many instances even with a lightweight harness, reducing the marginal benefit of additional harness mechanisms.

Table~\ref{tab:trajectory-stats-swe-deepseek} further shows that OpenCode changes DeepSeek execution behavior more aggressively than it does for Qwen. For DeepSeek-V3.1 and DeepSeek-V3.2, OpenCode leads to substantially more steps, tool calls, and prompt tokens than mini-SWE-agent, suggesting broader exploration and heavier use of the product-level workflow. However, for DeepSeek-V4-Pro, OpenCode uses fewer steps than mini-SWE-agent while still making more tool calls per step. This indicates that the strongest DeepSeek model can interact with OpenCode in a more compressed and tool-intensive manner, but the resulting performance gain over mini-SWE-agent becomes very small.

\begin{table}[t]
\centering
\caption{Average sample-level statistics of DeepSeek models.}
\label{tab:trajectory-stats-swe-deepseek}
\resizebox{\linewidth}{!}{\begin{tabular}{llcccc}
\toprule
Harness & Model & Step & Tool & Prompt & Response \\
\midrule
mini-SWE & DeepSeek-R1 & 17.71 & 15.97 & 127,552 & 31,280 \\
mini-SWE & DeepSeek-V3 & 15.90 & 14.69 & 104,289 & 1,886 \\
mini-SWE & DeepSeek-V3.1 & 58.66 & 56.66 & 1,000,415 & 6,905 \\
mini-SWE & DeepSeek-V3.2 & 79.99 & 72.23 & 1,687,660 & 13,665 \\
mini-SWE & DeepSeek-V4-Pro & 61.77 & 60.12 & 1,860,832 & 29,326 \\
\midrule
OpenCode & DeepSeek-R1 & 8.98 & 7.85 & 146,731 & 20,787 \\
OpenCode & DeepSeek-V3 & 36.42 & 34.66 & 1,163,976 & 5,488 \\
OpenCode & DeepSeek-V3.1 & 74.91 & 74.08 & 2,874,875 & 10,661 \\
OpenCode & DeepSeek-V3.2 & 89.42 & 101.18 & 2,796,537 & 19,093 \\
OpenCode & DeepSeek-V4-Pro & 31.96 & 70.42 & 1,178,990 & 16,397 \\
\bottomrule
\end{tabular}}
\end{table}

\finding{1.1}{
(1) \textbf{The marginal benefit of complex harnesses generally decreases as model capability improves}: on SWE-bench Verified, matched-model gains over mini-SWE-agent drop from +9.60/+13.60 pp to +4.07/+5.47 pp, and in Qwen SWE-bench Pro experiments the OpenCode--mini-SWE gap narrows from 6.00 pp to 2.67 pp. (2) \textbf{This convergence is not uniform across model families}. Qwen follows a smooth narrowing trend, whereas DeepSeek shows a non-monotonic pattern, with OpenCode helping more on intermediate models (e.g., DeepSeek-V3.1, DeepSeek-V3.2) and nearly converging with mini-SWE on DeepSeek-V4-Pro (34.3\% vs. 33.7\%).
}

\subsubsection{RQ1.2-Harness Performance on Different SE Tasks.}

To verify whether the findings from RQ1.1 are generally applicable, in RQ1.2 we select ProgramBench and GitTaskBench and conduct experiments using the same configuration as RQ1.1. We randomly sample 70 of the 200 instances from ProgramBench, including 20 easy, 21 medium, and 29 hard instances, and use all 54 instances from GitTaskBench.

\begin{figure}[!t]
\centering
\includegraphics[width=\linewidth]{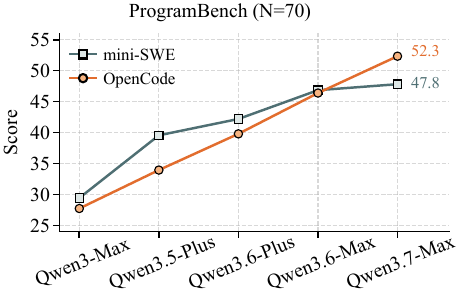}
\caption{Qwen-model scaling comparison between different harnesses on ProgramBench (N=70).}
\label{fig:programbench_70_qwen}
\end{figure}

\begin{figure}[!t]
\centering
\includegraphics[width=\linewidth]{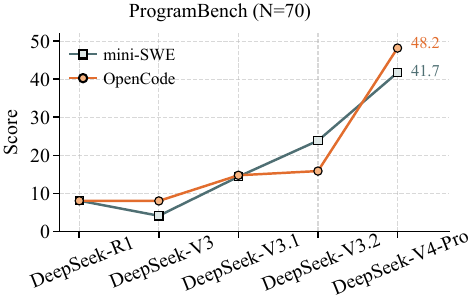}
\caption{DeepSeek-model scaling comparison between different harnesses on ProgramBench (N=70).}
\label{fig:programbench_70_ds}
\end{figure}

\begin{figure}[!t]
\centering
\includegraphics[width=\linewidth]{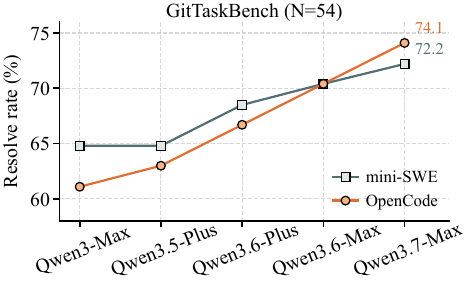}
\caption{Qwen-model scaling comparison between different harnesses on GitTaskBench (N=54).}
\label{fig:gittaskbench_54_qwen}
\end{figure}

\begin{figure}[!t]
\centering
\includegraphics[width=\linewidth]{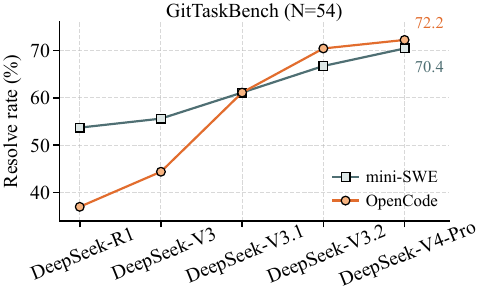}
\caption{DeepSeek-model scaling comparison between different harnesses on GitTaskBench (N=54).}
\label{fig:gittaskbench_54_ds}
\end{figure}

As shown in Figures~\ref{fig:programbench_70_qwen} and~\ref{fig:programbench_70_ds}, we find that on ProgramBench, OpenCode does not bring greater improvements to weaker models than mini-SWE-agent. Only the latest models (e.g., Qwen3.7-Max and DeepSeek-V4-Pro) can consistently leverage the complex harness to complete tasks more effectively. On GitTaskBench (Figures~\ref{fig:gittaskbench_54_qwen} and~\ref{fig:gittaskbench_54_ds}), we also observe a similar phenomenon: Qwen3.6-Max and DeepSeek-V3.1 are the respective turning points within their model families at which models begin to leverage a complex harness such as OpenCode.

The above results indicate that model capability is not the only factor affecting the effectiveness of a harness; task type and difficulty are also important factors. Compared with ProgramBench and GitTaskBench, SWE-bench Pro has a relatively fixed workflow, and code modifications typically involve only functions in a small number of files. In contrast, ProgramBench is a repository-generation task from scratch, often requiring the model to generate tens of thousands of lines of code while covering fine-grained functionality. The difficulty of GitTaskBench lies in its task flexibility: it involves multiple objectives and requires the model to independently understand the task goal, leverage repository information, and design an execution workflow. In addition, given the growing industry attention to SWE-style tasks in recent years, models may have received task-specific training, which could make them inherently stronger on this type of task and thereby reduce their dependence on the harness.

\finding{1.2}{
\textbf{The effectiveness of a harness is influenced not only by model capability, but also by task complexity and task type.} For SWE-bench Pro, where the workflow is relatively fixed, the amount of code modification is small, and the task has received sustained attention over time, models generally demonstrate strong adaptability. As model capability improves, the additional gains brought by a complex harness (e.g., OpenCode) gradually diminish. In contrast, for newer and more complex SE tasks such as ProgramBench and GitTaskBench, complex harnesses can provide greater benefits to stronger models (e.g., Qwen3.7-Max and DeepSeek-V4-Pro).
}

\subsection{RQ2: Component-Level Effect Analysis}

In RQ1, we treat the harness as a whole and analyze its performance across different SE tasks. Although the results show that complex harnesses (e.g., OpenCode) can improve performance compared with basic harnesses (e.g., mini-SWE-agent), the sources of these improvements remain unclear. Therefore, in RQ2, we decompose and combine representative components of the harness to construct the lightweight and modular \toolname{}. We then conduct ablation experiments on the full ProgramBench set with \toolname{}, using two advanced LLMs, Qwen3.7-Max and DeepSeek-V4-Pro.

\begin{figure}[!t]
\centering
\includegraphics[width=\linewidth]{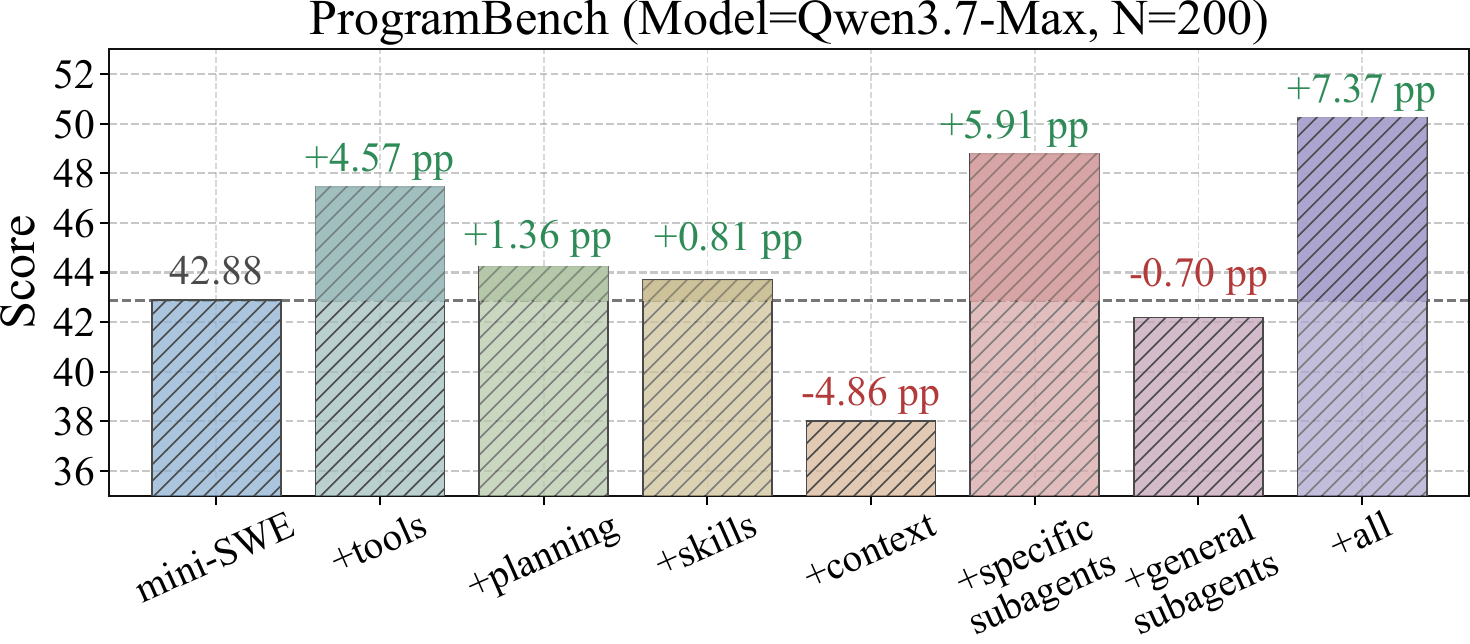}
\caption{Effect of different harness components on ProgramBench scores using Qwen3.7-Max.}
\label{fig:programbench_full_qwen}
\end{figure}

\begin{figure}[!t]
\centering
\includegraphics[width=\linewidth]{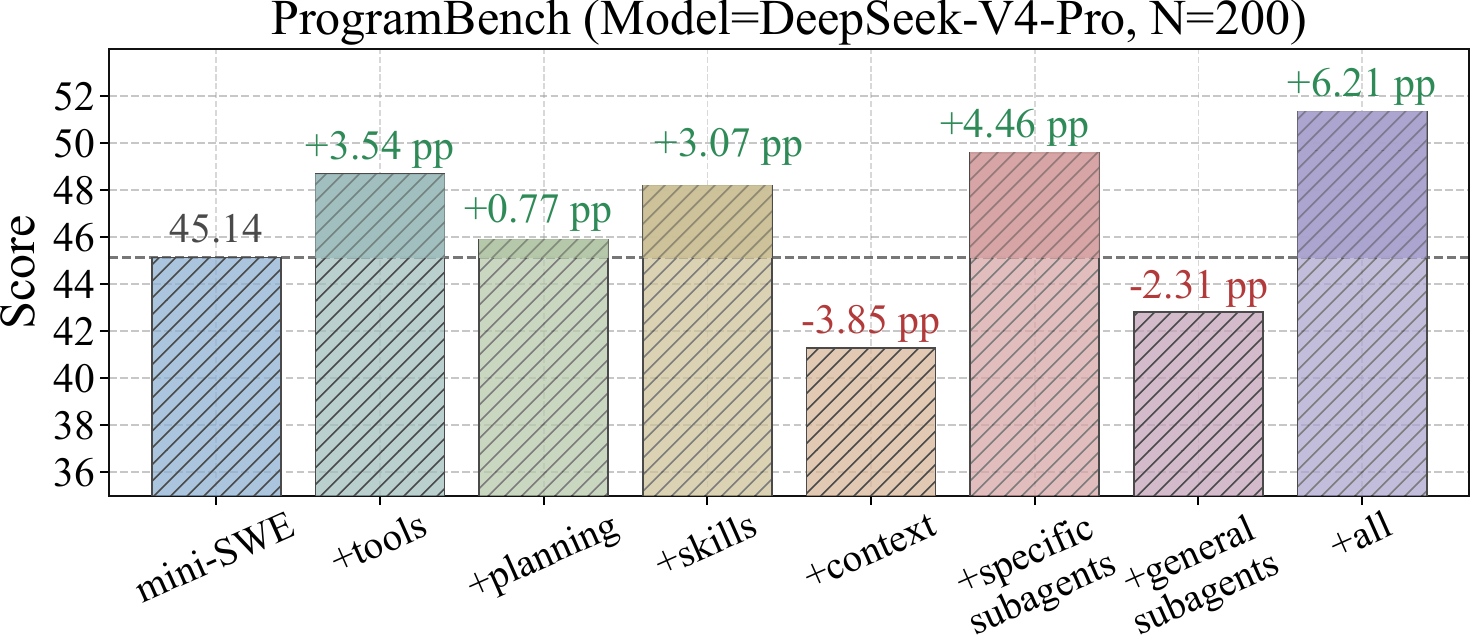}
\caption{Effect of different harness components on ProgramBench scores using DeepSeek-V4-Pro.}
\label{fig:programbench_full_ds}
\end{figure}

\subsubsection{RQ2.1-Individual Effect}

Figures~\ref{fig:programbench_full_qwen} and~\ref{fig:programbench_full_ds} show that different harness components contribute unevenly to agent performance. Among the individual components, the tool registry and task-specific subagents bring the most consistent improvements across both models. 
For Qwen3.7-Max, adding the tool registry improves the ProgramBench score by 4.57 percentage points, while task-specific subagents bring the largest single-component gain of 5.91 percentage points. 
For DeepSeek-V4-Pro, the same two components also provide strong gains of 3.54 and 4.46 points, respectively. These results suggest that structured repository operations and task-oriented delegation are particularly useful for repository-generation tasks, where agents must inspect files, create or modify multiple modules, and repeatedly validate implementation behavior.

Planning and lazy skills also improve performance, but their effects are more moderate and model-dependent. Explicit planning yields small gains for both Qwen3.7-Max and DeepSeek-V4-Pro, indicating that lightweight task lists can help organize long-horizon execution, but planning alone is insufficient to substantially improve repository generation. Lazy skills show a larger benefit for DeepSeek-V4-Pro than for Qwen3.7-Max, suggesting that procedural guidance is more helpful when the model can effectively decide when and how to invoke it.

In contrast, context compression and general subagents degrade performance for both models. Context compression reduces the score by 4.86 percentage points for Qwen3.7-Max and 3.85 percentage points for DeepSeek-V4-Pro, because ProgramBench requires preserving detailed requirements, interface constraints, and implementation states over long trajectories. Aggressive compression may remove information that is still useful for later coding and debugging. General subagents also underperform, indicating that generic SE delegation may introduce coordination overhead or produce artifacts that are not directly aligned with the fine-grained requirements of repository generation.

\finding{2.1}{Individual harness components have heterogeneous effects. Tool registry and task-specific subagents provide the most stable and substantial gains, planning and lazy skills offer smaller or model-dependent improvements, while context compression and general subagents can hurt performance on repository-generation tasks.}

\subsubsection{RQ2.2-Combined Effect}
When all components are combined, \toolname{} achieves the best performance among the variants for both models. For Qwen3.7-Max, the full configuration improves the score from 42.88 to 50.25, yielding a 7.37-point gain over mini-SWE-agent. For DeepSeek-V4-Pro, it improves the score from 45.14 to 51.35, corresponding to a 6.21-point gain. These improvements are larger than those of any single component, showing that multiple harness mechanisms can complement one another when integrated into a unified workflow.

However, the combined improvement is not simply the sum of all individual gains. Some components, such as context compression and general subagents, have negative effects in isolation, and their interaction with stronger components may introduce additional overhead. This indicates that harness components are not independent: their benefits depend on how they are coordinated within the agent loop. Nevertheless, the full \toolname{} configuration still consistently improves both models, suggesting that positive components such as structured tools, task-specific delegation, and procedural skills can dominate the overall effect when combined properly.

Compared with product-level harnesses, \toolname{} also achieves competitive performance while remaining lightweight and modular. On Qwen3.7-Max, \toolname{} reaches 50.25, reducing the gap to OpenCode (51.68) and Claude Code (52.33) to 1.43 and 2.08 percentage points, respectively. On DeepSeek-V4-Pro, \toolname{} achieves 51.35, which is also close to OpenCode (52.48) and Claude Code (52.76), with gaps of only 1.13 and 1.41 percentage points. These results show that although product-level harnesses still obtain the best scores, a lightweight modular harness composed of representative components can recover most of their performance gains. This suggests that the benefit of complex harnesses does not necessarily come from their full engineering stack; instead, a small set of carefully selected and coordinated components can account for a substantial portion of the improvement.

\finding{2.2}{The full \toolname{} configuration achieves the strongest performance among variants, improving over mini-SWE-agent on both models and outperforming any single component. Although \toolname{} still slightly trails product-level harnesses such as OpenCode and Claude Code, it closes most of the gap with a lightweight modular design, suggesting that a compact set of representative components can reproduce much of the benefit of complex harnesses.}

\begin{table}[t]
\centering
\caption{Comparing \toolname{} with other product-level harnesses on ProgramBench.}
\label{tab:comparison_with_harness_results}
\resizebox{\linewidth}{!}{%
\begin{tabular}{lcc}
\toprule
\textbf{Harness} & \textbf{Qwen3.7-Max} & \textbf{DeepSeek-V4-Pro} \\
\midrule
mini-SWE-agent & 42.88 & 45.14 \\
\toolname{}    & 50.25 \textcolor{green!50!black}{(+7.37\%)} &  51.35  \textcolor{green!50!black}{(+6.21\%)}  \\
OpenCode       & 51.68 \textcolor{green!50!black}{(+8.80\%)} &  52.48 \textcolor{green!50!black}{(+7.34\%)}   \\
ClaudeCode     & 52.33 \textcolor{green!50!black}{(+9.45\%)} &  52.76 \textcolor{green!50!black}{(+7.62\%)}   \\
\bottomrule
\end{tabular}}
\end{table}

\subsection{RQ3: Effects of Harness Mechanisms on Agent Behavior and Efficiency}

In RQ3, we collect statistics on the performance and cost of each configuration in RQ2, and further analyze execution trajectories to examine how harness components affect agent behavior. Specifically, RQ3.1 analyzes the differences in agent steps, tool calls, and token usage across different configurations of \toolname{}, while RQ3.2 measures the distribution patterns of probing behavior and uses case studies to illustrate shifts in failure modes.

\subsubsection{RQ3.1-Effects on Tool Use, Context Usage, and Token Usage}

Table~\ref{tab:programbench-cost} shows three main behavior patterns. First, tool-oriented mechanisms increase repository interaction: the tool registry modestly raises tool calls, and task-specific subagents increase tool calls by 115.76\% for Qwen3.7-Max and 73.10\% for DeepSeek-V4-Pro. Since task-specific subagents improve performance while general subagents do not, additional tool use appears beneficial only when it is directed toward the task. Second, context compression reduces cost but hurts performance. It cuts prompt tokens by 54.29\% and 77.02\% for the two models, yet RQ2 shows corresponding score drops of 4.86 and 3.85 percentage points, suggesting that compressed histories may lose requirements or debugging state needed later. Third, the full \toolname{} configuration improves performance with controlled prompt growth: although tool calls increase by 131.18\% and 83.58\%, prompt-token growth is much smaller (+16.37\% and +7.26\%), while scores improve by 7.37 and 6.21 percentage points over mini-SWE-agent. This suggests that the combined harness gains come from more structured interaction rather than simply spending more prompt budget.

\finding{3.1}{Harness mechanisms affect execution behavior in three main ways. Tools and subagents increase tool use, context compression reduces cost but may hurt performance, and the full \toolname{} configuration improves performance through more structured interaction with controlled prompt growth.}

\begin{table*}[t]
\centering
\caption{Average sample-level statistics of Qwen3.7-Max and DeepSeek-V4-Pro with different harness components on ProgramBench. Percentages in parentheses indicate relative changes compared with mini-SWE of the same model.}
\label{tab:programbench-cost}
\resizebox{0.95\linewidth}{!}{%
\begin{tabular}{llcccc}
\toprule
Model & Configuration & Step & Tool & Prompt & Response \\
\midrule
Qwen3.7-Max & mini-SWE & 104.13 & 104.14 & 3,978,148 & 46,763 \\
Qwen3.7-Max & + tools & 105.89 \textcolor{green!50!black}{(+1.69\%)} & 120.87 \textcolor{green!50!black}{(+16.06\%)} & 4,299,572 \textcolor{green!50!black}{(+8.08\%)} & 44,366 \textcolor{red}{(-5.13\%)} \\
Qwen3.7-Max & + planning & 102.19 \textcolor{red}{(-1.86\%)} & 104.35 \textcolor{green!50!black}{(+0.20\%)} & 4,817,378 \textcolor{green!50!black}{(+21.10\%)} & 65,679 \textcolor{green!50!black}{(+40.45\%)} \\
Qwen3.7-Max & + skills & 108.69 \textcolor{green!50!black}{(+4.38\%)} & 109.20 \textcolor{green!50!black}{(+4.86\%)} & 4,503,568 \textcolor{green!50!black}{(+13.21\%)} & 47,913 \textcolor{green!50!black}{(+2.46\%)} \\
Qwen3.7-Max & + context & 61.77 \textcolor{red}{(-40.68\%)} & 62.06 \textcolor{red}{(-40.41\%)} & 1,818,344 \textcolor{red}{(-54.29\%)} & 26,577 \textcolor{red}{(-43.17\%)} \\
Qwen3.7-Max & + specific subagents & 141.29 \textcolor{green!50!black}{(+35.69\%)} & 224.69 \textcolor{green!50!black}{(+115.76\%)} & 4,270,462 \textcolor{green!50!black}{(+7.35\%)} & 72,240 \textcolor{green!50!black}{(+54.48\%)} \\
Qwen3.7-Max & + general subagents & 99.06 \textcolor{red}{(-4.87\%)} & 157.40 \textcolor{green!50!black}{(+51.14\%)} & 4,372,546 \textcolor{green!50!black}{(+9.91\%)} & 67,919 \textcolor{green!50!black}{(+45.24\%)} \\
Qwen3.7-Max & + all (\toolname{}) & 133.68 \textcolor{green!50!black}{(+28.38\%)} & 240.75 \textcolor{green!50!black}{(+131.18\%)} & 4,629,172 \textcolor{green!50!black}{(+16.37\%)} & 64,751 \textcolor{green!50!black}{(+38.47\%)} \\
\midrule
DeepSeek-V4-Pro & mini-SWE & 142.19 & 181.84 & 10,065,204 & 91,370 \\
DeepSeek-V4-Pro & + tools & 164.11 \textcolor{green!50!black}{(+15.41\%)} & 204.20 \textcolor{green!50!black}{(+12.29\%)} & 11,979,965 \textcolor{green!50!black}{(+19.02\%)} & 86,982 \textcolor{red}{(-4.80\%)} \\
DeepSeek-V4-Pro & + planning & 140.66 \textcolor{red}{(-1.07\%)} & 184.27 \textcolor{green!50!black}{(+1.33\%)} & 10,418,872 \textcolor{green!50!black}{(+3.51\%)} & 91,832 \textcolor{green!50!black}{(+0.51\%)} \\
DeepSeek-V4-Pro & + skills & 147.85 \textcolor{green!50!black}{(+3.98\%)} & 179.22 \textcolor{red}{(-1.44\%)} & 10,415,011 \textcolor{green!50!black}{(+3.48\%)} & 90,942 \textcolor{red}{(-0.47\%)} \\
DeepSeek-V4-Pro & + context & 54.17 \textcolor{red}{(-61.91\%)} & 56.61 \textcolor{red}{(-68.87\%)} & 2,312,701 \textcolor{red}{(-77.02\%)} & 33,747 \textcolor{red}{(-63.07\%)} \\
DeepSeek-V4-Pro & + specific subagents & 190.45 \textcolor{green!50!black}{(+33.94\%)} & 314.77 \textcolor{green!50!black}{(+73.10\%)} & 9,104,564 \textcolor{red}{(-9.54\%)} & 118,453 \textcolor{green!50!black}{(+29.64\%)} \\
DeepSeek-V4-Pro & + general subagents & 169.54 \textcolor{green!50!black}{(+19.24\%)} & 332.14 \textcolor{green!50!black}{(+82.65\%)} & 15,284,596 \textcolor{green!50!black}{(+51.86\%)} & 145,124 \textcolor{green!50!black}{(+58.83\%)} \\
DeepSeek-V4-Pro & + all (\toolname{}) & 203.53 \textcolor{green!50!black}{(+43.14\%)} & 333.84 \textcolor{green!50!black}{(+83.58\%)} & 10,795,507 \textcolor{green!50!black}{(+7.26\%)} & 103,443 \textcolor{green!50!black}{(+13.21\%)} \\
\bottomrule
\end{tabular}}
\end{table*} 

\subsubsection{RQ3.2-Effects on Failure Modes} 
\begin{figure}[!t]
\centering
\includegraphics[width=\linewidth]{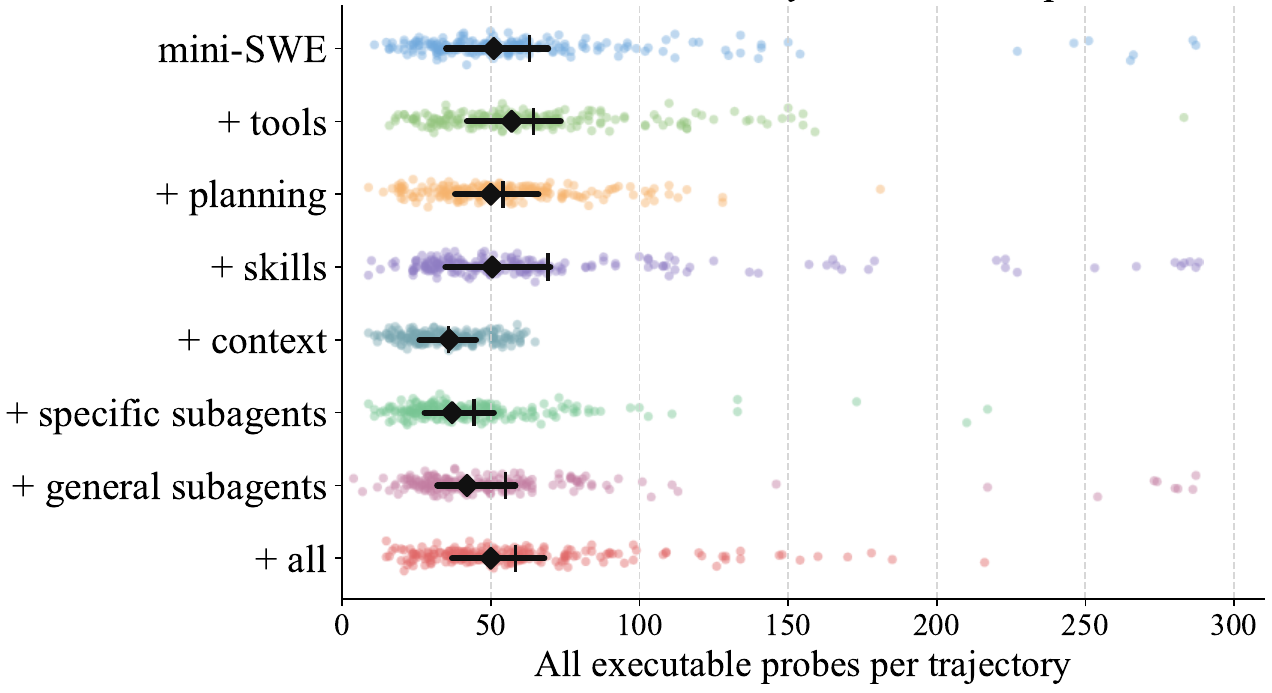}
\caption{Distribution of probing times (Qwen3.7-Max).}
\label{fig:qwen_probe_distribution}
\end{figure}

Through manual inspection of failed and low-scoring trajectories produced by mini-SWE-agent, we categorize failures by probing count and validation behavior and find that models mainly exhibit two failure modes on ProgramBench. (1) \textbf{Excessive probing.} At the early exploration stage, the agent blindly performs a large number of fine-grained behavioral probes on the reference binary, but fails to cover the core functionality of the target program. As a result, it exhausts the budget and finishes the task hastily. (2) \textbf{Insufficient probing.} The agent assumes that it has already understood the core functionality of the program after only a small number of probes (e.g., $\leq 20$), and after implementing the program, it validates the solution using only a few smoke tests. As shown in Figure~\ref{fig:qwen_probe_distribution}, we compare the distribution of probing times for Qwen3.7-Max under different harness configurations. We observe that the components with the largest performance gains, i.e., task-specific subagents, substantially reduce the tendency of mini-SWE-agent to perform either excessive or insufficient probing. Building on these improvements, \toolname{} further combines the strengths of different components. It not only reduces extreme probing behaviors, but also learns to use more probes to explore the target program behavior more comprehensively, leading to a stable rightward shift in the overall distribution of probing counts.

Two cases illustrate how \toolname{} mitigates these extremes. For excessive probing,
\texttt{jqlang\_\_jq.b33a763} shows how \toolname{} reduces unproductive
exploration: mini-SWE-agent issued 133 reference-binary probes and passed only
1602/6491 tests (24.68\%), whereas \toolname{} reduced the total number of
reference probes to 43 and improved the result to 4425/6491 tests (68.17\%).
For insufficient probing, \texttt{lfos\_\_calcurse.49180d5} shows the opposite
correction: mini-SWE-agent made only 3 reference probes before implementation and
passed 1231/1994 tests (61.74\%), while \toolname{} expanded exploration to 45
reference probes and improved the score to 1499/1994 tests (75.18\%). These
examples suggest that \toolname{} does not simply increase or decrease probing
uniformly; instead, it regularizes exploration by reducing wasteful probing when
the agent is stuck and increasing behavioral coverage when the agent is
prematurely confident.

\finding{3.2}{Harness mechanisms affect failure modes by regularizing probing behavior. On ProgramBench, effective components such as task-specific subagents reduce both excessive low-value probes and insufficient shallow validation. By combining these components, \toolname{} enables more stable behavioral exploration and helps avoid wasteful probing and premature implementation.}

\section{Related Work}

\subsection{LLM-based Agents for Software Engineering}

LLM-based agents have been widely studied for code generation~\cite{liu2023your,zhuo2025bigcodebench,jain2025livecodebench}, repository-level problem solving~\cite{jimenez2024swe,zan2026multi}, and multi-role software development workflows such as ChatDev~\cite{qian2024chatdev} and MetaGPT~\cite{hong2024metagpt}. Other work improves agentic coding through planning, retrieval, debugging, and tool feedback~\cite{islam2024mapcoder,liu2024repobench,li2024deveval}. For repair and issue resolution, SWE-bench~\cite{jimenez2024swe} has motivated systems such as SWE-agent~\cite{yang2024swe}, AutoCodeRover~\cite{zhang2024autocoderover}, RepairAgent~\cite{bouzenia2025repairagent}, and Agentless~\cite{xia2024agentless}, while newer benchmarks extend evaluation to broader repository-level tasks~\cite{yang2026programbench,ni2026gittaskbench}.

These studies demonstrate the value of agentic workflows, but they usually evaluate complete systems, benchmarks, or task-specific pipelines. As a result, the effects of the model, harness, and task are often entangled. Our work complements them by treating harness design as the main research object and measuring how harnesses and their components affect SE-agent performance across models and tasks.

\subsection{Agent Harness}

Recent studies emphasize that agent performance depends not only on the base model, but also on the surrounding harness or external scaffolding~\cite{ning2026code,pan2026natural,meng2026agent,zhou2026externalization,lin2026agentic}. General agent frameworks such as ReAct~\cite{yao2022react}, ToolLLM~\cite{qin2024toolllm}, Reflexion~\cite{shinn2023reflexion}, Voyager~\cite{wang2023voyager}, and AutoGen~\cite{wu2024autogen} show that tool use, memory, planning, reflection, skills, and multi-agent coordination can improve long-horizon problem solving.

In SE, harnesses must support repository navigation, editing, command execution, testing, and recovery. Existing systems introduce mechanisms such as agent-computer interfaces~\cite{yang2024swe}, repository search and fault localization~\cite{zhang2024autocoderover,bouzenia2025repairagent}, and localization-repair-validation pipelines~\cite{xia2024agentless}. However, these mechanisms are usually evaluated as parts of complete agents. Our work isolates representative harness components and analyzes their performance, behavioral, and efficiency effects under controlled model and task settings.

\section{Threats to Validity}
\textbf{Internal Threat.}
The main internal threat stems from our limited selection of harness components: we study five representative components, which may omit other relevant mechanisms. We plan to extend the evaluation to more components in future work.
A second threat arises from the coverage of model families, as we use only Qwen and DeepSeek.
However, this is mitigated by covering 10 most representative variants, with total prompt-side token usage exceeding \textbf{10B} (Qwen) and \textbf{20B} (DeepSeek) at an estimated cost of over \textbf{\$7,000} USD.

\textbf{External Threat.} The main external threat stems from our dataset selection. Our primary experiments are conducted on ProgramBench, so the conclusions may not directly generalize to other datasets. Nevertheless, ProgramBench represents a recent and challenging benchmark for repository-level software generation, covering the complete development lifecycle from requirement understanding to artifact generation. Therefore, we believe ProgramBench provides a meaningful and comprehensive testbed for evaluating the studied harnesses.

\section{Conclusion}
This paper studies harness design as a first-class factor in LLM-based SE agents. Across three benchmarks, we show that harness gains depend on model capability and task type. Through \toolname{}, we further identify which components matter: structured tools and task-specific subagents improve performance, while context compression and general subagents can hurt. Our trajectory analysis shows that effective harnesses regularize exploration and tool use, suggesting that future SE-agent design should jointly consider models, harnesses, tasks, and execution behavior.

\footnotesize
\bibliographystyle{plainnat}
\bibliography{reference}

\end{document}